\documentclass[aps,prd,letterpaper,showpacs,preprintnumbers,twocolumn,amsmath,amssymb,nofootinbib]{revtex4}

\DeclareMathAlphabet   {\mathsc}{OT1}{cmr}{m}{sc}

\def\[{\left [}
\def\]{\right ]}
\def\({\left (}
\def\){\right )}
\newcommand{\lang}{\left\langle}
\newcommand{\rang}{\right\rangle}

\newcommand{\oline}[1]{\overline{#1}}

\newcommand{\Lag}{\mathcal{L}}

\newcommand{\wh}[1]{\widehat{#1}}

\newcommand{\GeV}      {~\mathrm{GeV}}

\newcommand{\PL}       {\mathsc{pl}}

\newcommand{\hc}       {\mathrm{\; h.c. \;}}

\newcommand{\Tr}{{\rm Tr}}

\newcommand{\gappeq}{\mathrel{\rlap {\raise.5ex\hbox{$>$}}
{\lower.5ex\hbox{$\sim$}}}}
\newcommand{\lappeq}{\mathrel{\rlap{\raise.5ex\hbox{$<$}}
{\lower.5ex\hbox{$\sim$}}}}

\begin{document}

\preprint{UPR-1126T}

\title{String-Inspired Triplet See-Saw from Diagonal Embedding of
$SU(2)_L \subset SU(2)_A \times SU(2)_B$}

\author{Paul~Langacker}
\author{Brent D.~Nelson} 
\affiliation{Department of Physics \& Astronomy, University of
Pennsylvania, Philadelphia, PA 19104, USA}

\date{\today}

\begin{abstract}
Motivated by string constructions, we consider a variant on the
Type~II see-saw mechanism involving the exchange of triplet
representations of $SU(2)_L$ in which this group arises from a
diagonal embedding into $SU(2)_A \times SU(2)_B$. A natural
assignment of Standard Model lepton doublets to the two underlying
gauge groups results in a bimaximal pattern of neutrino mixings
and an inverted hierarchy in masses. Simple perturbations around
this leading-order structure can accommodate the observed pattern
of neutrino masses and mixings.
\end{abstract}


\maketitle

\section*{Introduction}

Observations by a variety of experimental collaborations have now
firmly established the hypothesis that neutrino oscillations occur
and that they are the result of non-vanishing neutrino masses and
mixing angles~\cite{Maltoni:2003da,Bahcall:2004ut,Fogli:2005cq}.
While our knowledge of neutrino mass-differences and mixings has
continued to improve over recent years, there continues to be as
yet no consensus on the correct mechanism for generating the quite
small neutrino masses implied by the experimental data. In some
respects this is similar to the case of quark masses and mixings:
despite having access to even more of the relevant experimental
data for an even longer period of time, no compelling model of the
hierarchies of masses and mixings in the quark sector has emerged
either. But most theoretical effort in the area of neutrinos goes
beyond the simple Dirac-mass Yukawa operator by introducing new
structures in the superpotential to account for neutrino masses,
such as the see-saw mechanism (in one of its various forms, to be
defined more precisely below).\footnote{For some recent reviews of
theoretical models of neutrino masses and mixings,
see~\cite{King:2003jb,DeGouvea:2005gd,Langacker:2004xy,Altarelli:2004za,Mohapatra:2004vr}
and references therein.} Thus neutrinos are likely to be very
special in the Standard Model -- and its supersymmetric extensions
-- and may thereby provide a unique window into high-scale
theories that the quark sector fails to illuminate.

It has thus far been mostly in vain that we might look to string
theory for some guidance in how to approach the issue of flavor.
In part this is because of the vast number of possible vacua in
any particular construction, each with its own set of fields and
superpotential couplings between them. On the other hand the
problem of generating small neutrino masses may be one of the most
powerful discriminants in finding realistic constructions. This
was one of the conclusions of a recently completed
survey~\cite{Giedt:2005vx} of a large class of explicit orbifold
compactifications of the heterotic string for the standard (or
``Type~I'') see-saw in its minimal form. The fact that no such
viable mechanism was found may suggest that often-neglected
alternatives to the standard see-saw may have more theoretical
motivation than considerations of simplicity, elegance, or GUT
structure would otherwise indicate.

In this work we study the properties of a new construction of
see-saw mechanisms that is motivated by known string
constructions. The mechanism is an example of the Higgs triplet or
``Type II''
seesaw~\cite{Lazarides:1980nt,Mohapatra:1980yp,Schechter:1981cv,Ma:1998dx,Hambye:2000ui,Rossi:2002zb},
but the stringy origin has important implications for the mixings
and mass hierarchy that distinguish it from conventional
``bottom-up'' versions of the triplet model. After outlining the
model in a general way below we will motivate its plausibility in
string theory by considering a particular $\mathbb{Z}_3 \times
\mathbb{Z}_3$ orbifold of the heterotic string~\cite{Font:1989aj},
where several of the properties needed for a fully realistic model
are manifest.

\section{General Features of $SU(2)$ Triplet Models}
\label{sec:general}

Let us briefly review the form of the effective neutrino mass
matrix to establish our notation and to allow the contrast between
models involving triplets of $SU(2)_L$ and those involving
singlets to be more apparent. While models of neutrino masses can
certainly be considered without low-energy supersymmetry, our
interest in effective Lagrangians deriving from string theories
which preserve $N=1$ supersymmetry leads us to couch our
discussion in a supersymmetric framework. Then the effective mass
operator involving only the light (left-handed) neutrinos has mass
dimension five. Once the Higgs fields acquire vacuum expectation
values (vevs) the effective neutrino masses are given by
\begin{equation}
(\mathbf{m}_{\nu})_{ij} = (\lambda_{\nu})_{ij} \frac{v_2^2}{M} ,
\label{mnu} \end{equation}
where $v_2$ is the vev of the Higgs doublet $H_2$ with hypercharge
$Y=+1/2$. The $3 \times 3$ matrix of couplings $\lambda_{\nu}$ is
necessarily symmetric in its generation indices~$i$ and~$j$.

Such an operator can be induced through the exchange of heavy
singlet (right-handed) neutrinos $N_R$ -- as in the standard or
``Type~I'' see-saw approach~\cite{GRS,Yanagida,Valle1} -- or
through the exchange of heavy triplet states
$T$~\cite{Lazarides:1980nt,Mohapatra:1980yp,Schechter:1981cv,Ma:1998dx,Hambye:2000ui,Rossi:2002zb},
or both. In either case, the mass scale $M$ is given by the scale
at which lepton number is broken (presumably the mass scale of the
heavy state being exchanged). In the presence of both
contributions to the light neutrino masses we have the general
mass matrix
\begin{equation}
\Lag = \frac{1}{2} \( \oline{\nu}_L \; \; \oline{N}_L^c \) \(
\begin{array}{cc} \mathbf{m}_{T} & \mathbf{m}_{D} \\ \mathbf{m}_{D}^{T} &
\mathbf{m}_{S} \end{array} \) \( \begin{array}{c} \nu_R^c \\ N_R
\end{array} \) + \hc .
\label{general} \end{equation}
Each of the four quantities in~(\ref{general}) are understood to
be $3 \times 3$ matrices in flavor space. That is, we imagine a
model with one {\em species} of lepton doublet $L$ (with three
generations) and, if present, one {\em species} of right-handed
neutrino field (with three generations).

We will use the name ``triplet models'' to refer to any model
which uses electroweak triplet states alone to generate neutrino
masses. That is, such a model dispenses with right-handed
neutrinos altogether, and the effective neutrino mass
in~(\ref{mnu}) is then simply identified with the entry
$\mathbf{m}_{T}$ in~(\ref{general}). A supersymmetric extension of
the MSSM capable of giving small effective masses to left-handed
neutrinos would involve two new sets of fields $T_i$ and
$\oline{T}_i$ which transform as triplets under $SU(2)_{L}$ and
have hypercharge assignments $Y_{T} = +1$ and $Y_{\oline{T}} =
-1$, respectively~\cite{Hambye:2000ui,Rossi:2002zb}. For the time
being we will consider just one pair of such fields, which couple
to the Standard Model through the superpotential
\begin{eqnarray}
W_{\nu} &=& (\lambda_{T})_{ij} L_i T L_j + \lambda_1 H_1 T H_1 +
\lambda_2 H_2 \oline{T} H_2  \nonumber \\
 & & + M_T T \oline{T} + \mu H_1 H_2 ,
\label{Wnu} \end{eqnarray}
where the $SU(2)$ indices on the doublets and triplet have been
suppressed. Strictly speaking, the coupling $\lambda_1$ is not
necessary to generate the required neutrino masses, but given the
Standard Model charge assignments of the fields $T$ and
$\oline{T}$ there is no {\em a priori} reason to exclude this
coupling. The mass scale $M$ in~(\ref{mnu}) is to be identified
with $M_T$ in this case, and the matrix $\lambda_{T}$ is symmetric
in its generation indices. From the Lagrangian determined
by~(\ref{Wnu}), it is clear that should the auxiliary fields of
the chiral supermultiplets for the triplets vanish in the vacuum
$\lang F^T \rang = \lang F^{\oline{T}} \rang = 0$, and we assume
no vevs for the left-handed sneutrino fields, then there is a
simple solution for the vevs of the neutral components of the
triplet fields
\begin{equation}
\lang T \rang = -\frac{\lambda_2 \lang H_2 \rang ^2}{M_T} \; ;
\quad \lang \oline{T} \rang = -\frac{\lambda_1 \lang H_1 \rang
^2}{M_T} ,
\label{Tvevs} \end{equation}
implying $m_{\nu} = \lambda_T \lang T \rang$.

Recent models of the Type~II
variety~\cite{King:2003jb,DeGouvea:2005gd,Langacker:2004xy,Altarelli:2004za,Mohapatra:2004vr}
would typically retain the right-handed neutrinos and utilize all
the components in the mass matrix of~(\ref{general}) to explain
the neutrino masses and mixings. These examples are often inspired
by SO(10) GUT considerations or are couched in terms of left-right
symmetry more generally. The latter commonly employ additional
Higgs fields transforming as $(1,\mathbf{3})$ under $SU(2)_L
\times SU(2)_R$, which acquire vevs to break the gauge group to
the Standard Model.

Instead we imagine a process by which the $SU(2)$ of the Standard
Model emerges as the result of a breaking to a {\em diagonal}
subgroup $SU(2)_A \times SU(2)_B \to SU(2)_L$ at a very high
energy scale. Furthermore, while we will employ two conjugate
triplet representations which form a vector-like pair under the
Standard Model gauge group with $Y = \pm 1$, we do not seek to
embed this structure into a left-right symmetric model.

\section{Diagonal Embedding of $SU(2)_L$}
\label{sec:diagonal}

In attempting to embed the framework of the previous section in a
model of the weakly coupled heterotic string we immediately
encounter an obstacle: the simplest string constructions contain
in their massless spectra only chiral superfields which transform
as fundamentals or (anti-fundamentals) of the non-Abelian gauge
groups of the low-energy theory. Scalars transforming as triplets
of $SU(2)$ simply do not exist for such affine level one
constructions~\cite{Gross:1985fr,Font:1990uw}. Indeed, scalars
transforming in the adjoint representation appear only at affine
level two, while representations such as the $\mathbf{120}$ and
$\mathbf{210}$ of $SO(10)$ appear only at affine level four -  and
the $\mathbf{126}$ of $SO(10)$, which contains triplets of the
Standard Model $SU(2)_L$, has been shown to never appear in
free-field heterotic string constructions~\cite{Dienes:1996yh}.

Directly constructing four-dimensional string compactifications
yielding higher affine-level gauge groups has proven to be a
difficult task. But a group factor $\mathcal{G}$ can be
effectively realized at affine level $k = n$ by simply requiring
it to be the result of a breaking of $\mathcal{G}_1 \times
\mathcal{G}_2 \times \cdots \times \mathcal{G}_n$ to the overall
diagonal subgroup. In fact, these two ways of understanding higher
affine levels -- picking a particular set of GSO projections in
the underlying string construction and the low energy field theory
picture of breaking to a diagonal subgroup -- are equivalent
pictures~\cite{Dienes:1996yh}. With this as motivation, let us
consider an appropriate variation on the superpotential
of~(\ref{Wnu}). The breaking of the gauge group $SU(2)_A \times
SU(2)_B$ to the diagonal subgroup, which we identify as $SU(2)_L$,
can occur through the vacuum expectation value of a field in the
bifundamental representation of the underlying product group via
an appropriately arranged scalar potential. For the purposes of
our discussion here we will need only assume that this breaking
takes place at a sufficiently high scale, say just below the
string compactification scale. As such ideas for product group
breaking have been considered in the past~\cite{Barbieri:1994jq},
we will not concern ourselves further with this step.

Any additional bifundamental representations will decompose into
triplet and singlet representations under the surviving $SU(2)_L$.
Gauge invariance of the underlying $SU(2)_A \times SU(2)_B$ theory
then {\em requires} that the neutrino-mass generating
superpotential coupling involving lepton doublets and our $SU(2)$
triplet now be given by
\begin{equation}
W_{\nu} = (\lambda_{T})_{ij} L_i T L'_j ,
\label{WT} \end{equation}
where the field $T$ is the $SU(2)_L$ triplet representation and
$L$ and $L'$ are two {\em different species} of doublets under the
$SU(2)_L$ subgroup. That is, we denote by $L$ fields which have
representation $(\mathbf{2},1)$ and by $L'$ fields which have
representation $(1, \mathbf{2})$ under the original $SU(2)_A
\times SU(2)_B$ gauge theory. These two sets of lepton doublets,
each of which may carry a generation index as determined by the
string construction, arise from different sectors of the string
Hilbert space. Once the gauge group is broken to the diagonal
subgroup this distinction between the species is lost {\em except
for the pattern of couplings represented by the matrix
$\mathbf{\lambda}_{T}$.} The indices $i$ and $j$ carried by the
lepton doublets represent internal degeneracies arising from the
specific construction. It is natural to identify these indices
with the flavor of the charged lepton (up to mixing effects, which
we assume to be small).

In a minimal model, with only three lepton doublets charged under
the Standard Model $SU(2)_L$, we are obliged to separate the
generations, with two arising from one sector of the theory and
one from the other. The precise form of the effective neutrino
mass matrix will depend on this model-dependent identification,
but one property is immediately clear: {\em the effective neutrino
mass matrix will necessarily be off-diagonal in the charged lepton
flavor-basis.}

We will restrict our study to the case of one triplet state with
supersymmetric mass $M_T$ as in~(\ref{Wnu}). If we separate the
doublet containing the electron from the other two, by defining
$L_i = L_e = (\mathbf{2},1)$ and $L'_j = L_{\mu}, L_{\tau} = (1,
\mathbf{2})$ under $SU(2)_A \times SU(2)_B$, then the matrix of
couplings $\lambda_T$ is (to leading order)
\begin{equation}
\lambda_T = \lambda_0 \(\begin{array}{ccc} 0 & a & b \\ a & 0 & 0
\\ b & 0 & 0 \end{array} \) .
\label{lambdaT1} \end{equation}
It is natural to assume that the overall coefficient $\lambda_0$
in~(\ref{lambdaT1}) is of order unity. In fact if we now return to
a string theory context, particularly that of the heterotic string
with orbifold compactification, then the fact that the two
generations of $L'_j$ in~(\ref{WT}) arise from the same sector of
the string Hilbert space (i.e., the same fixed point location
under the orbifold action) suggests that we should identify the
coupling strengths: $a=b$.

Neutrino mass matrices based on the texture in~(\ref{lambdaT1})
with $a=b$ are not new to this work, but were in fact considered
not long ago as a starting point for the bimaximal mixing
scenario~\cite{Barger:1998ta,Barbieri:1998mq,Altarelli:1998nx,Jezabek:1998du}.
In fact, the form of~(\ref{lambdaT1}) can be derived from a
bottom-up point of view by first postulating a new symmetry based
on the modified lepton number combination $L_e - L_{\mu} -
L_{\tau}$~\cite{Mohapatra:1999zr,Babu:2002ex}. Indeed, the
operator in~(\ref{WT}) with the identification of $L = L_e$ and
$L' = L_{\mu}, L_{\tau}$ does indeed conserve this quantum number.
However, in the string-theory motivated (top down) approach this
conserved quantity arises as an {\em accidental} symmetry
pertaining to the underlying geometry of the string
compactification. It reflects the different geometrical location
of the fields (in terms of orbifold fixed points) of the electron
doublet from the muon and tau doublets.

To make contact with data it is necessary to consider the Yukawa
interactions of the charged leptons as well. To that end, our
string-inspired model should have a superpotential of the form
\begin{eqnarray}
W &=& \lambda_T L T L' + \lambda_1 H_1 T H'_1 + \lambda_2 H_2
\oline{T} H'_2 \nonumber \\
 & & + \lambda_3 S_3 T \oline{T} + \lambda_4 S_4 H_1 H_2 +
 \lambda_5 S_5 H'_1 H'_2 \nonumber \\
 & & + \wh{\lambda}_4 \wh{S}_4 H_1 H'_2 +
 \wh{\lambda}_5 \wh{S}_5 H'_1 H_2 \nonumber \\
 & & + \lambda_6 S_6 L H_1 + \lambda_7 S_7 L' H'_1 \nonumber \\
 & & + \wh{\lambda}_6 \wh{S}_6 L H'_1 + \wh{\lambda}_7 \wh{S}_7 L' H_1 ,
\label{Wfull} \end{eqnarray}
where generation indices have been suppressed. The terms
proportional to the couplings $\lambda_1$ and $\lambda_2$
in~(\ref{Wnu}) must now be modified to reflect the fact that the
Higgs doublets must also come from two different species. These
are denoted in the same manner as the lepton doublets: $H_{1,2}$
for $(\mathbf{2},1)$ representations and $H'_{1,2}$ for
$(1,\mathbf{2})$ representations. The second line in~(\ref{Wfull})
are the dynamically-generated supersymmetric mass terms, with
$\lambda_3 \lang S_3 \rang \equiv M_T$. The fields $S_3$, $S_4$
and $S_5$ are singlets under $SU(2)_A \times SU(2)_B$ with
hypercharge $Y=0$, while $\wh{S}_4$ and $\wh{S}_5$ are $SU(2)_L$
singlets with $Y=0$ transforming as $(\mathbf{2},\mathbf{2})$
under $SU(2)_A \times SU(2)_B$. We anticipate a large vev for
$S_3$. The fields $S_4, S_5, \wh{S}_4$, and $\wh{S}_5$ may acquire
TeV scale vevs from supersymmetry breaking, leading to generalized
$\mu$ terms\footnote{The $\mu$ parameters of the Higgs scalar
potential could also arise as effective parameters only after SUSY
breaking via the Giudice-Masiero
mechanism~\cite{Giudice:1988yz}.}, or some could have vevs near
the string scale (or at an intermediate scale), projecting some of
the Higgs states out of the low energy theory. From the point of
view of both $SU(2)_L$ as well as the underlying $SU(2)_A \times
SU(2)_B$ it is not necessary that $S_4$ and $S_5$ be distinct
fields; there may be string selection rules forbidding their
identification in an explicit construction, however. Similar
statements apply to $\wh{S}_4$ and $\wh{S}_5$. Of course, some of
these fields could be absent.

The final line of~(\ref{Wfull}) represents the Dirac mass
couplings of the left-handed leptons with their right-handed
counterparts. Again, the fields $S_6$ and $S_7$, both singlets
under $SU(2)_A \times SU(2)_B$, carrying only hypercharge $Y=+1$,
may or may not be identified depending on the construction, while
$\wh{S}_6$ and $\wh{S}_7$, which may or may not be distinct,
transform as $(\mathbf{2},\mathbf{2})$. Some of these fields may
be absent. We assume that $S_6$, $S_7$, $\wh{S}_6$, and $\wh{S}_7$
do not acquire vevs. Charged lepton masses are then determined by
some combination of the coupling matrices $\lambda_6$,
$\lambda_7$, $\wh{\lambda}_6$ and $\wh{\lambda}_7$ (and possibly
higher-order terms that connect the two sectors) as well as
appropriate choices of Higgs vevs for the neutral components of
the four Higgs species.

\section{Making Contact with Experimental Data}
\label{sec:data}

Having laid out the framework for our string-based model, we now
wish to ask how well such a structure can accommodate the
measurements of neutrino mixing angles and mass differences that
have been made, and what sort of predictions (if any) might this
framework make in terms of future experimental observations. We
use a convention in which the solar mixing data defines the mass
difference between $m_2$ and $m_1$ with $m_2 > m_1$. Then the
eigenvalue $m_3$ relevant for the atmospheric data is the
``isolated'' eigenvalue.

The current experimental picture is summarized by the recent three
neutrino global analysis in~\cite{Fogli:2005cq}. For the solar
neutrino sector we take
\begin{eqnarray}
\Delta m_{12}^{2} & = & 7.92 (1\pm 0.09) \; \times 10^{-5} {\rm
eV}^2 \label{dm12} \\
\sin^{2}\theta_{12} & = & 0.314(1_{-0.15}^{+0.18}) ,
\label{theta12}
\end{eqnarray}
where all measurements are $\pm 2 \sigma$ (95\% C.L.). The last
measurement implies a value for the mixing angle $\theta_{12}$
itself of $\theta_{12} \simeq 0.595_{-0.052}^{+0.060}$, well below
the maximal mixing value $\theta_{12}^{max}=\pi/4$. We take the
upper bound on $\theta_{13}$ to be
\begin{equation}
\sin^{2} \theta_{13} =0.9_{-0.9}^{+2.3} \times 10^{-2} \; \quad
\Rightarrow |\theta_{13}| < 0.18
\label{theta13} \end{equation}
at the~2$\sigma$ level. For the atmospheric oscillations
\begin{eqnarray}
|\Delta m^2_{23}|& =& 2.4 (1_{-0.26}^{+0.21}) \times 10^{-3} {\rm
eV}^{2} \label{dm23} \\
\sin^{2}\theta_{23}& =& 0.44 (1_{-0.22}^{+0.41}),
\label{theta23} \end{eqnarray}
consistent with maximal mixing ($\sin^2 \theta_{23}^{\rm
max}=0.5$).

With this in mind, let us consider the general off-diagonal
Majorana mass matrix
\begin{equation}
m_{\nu} = \( \begin{array}{ccc} 0 & a & b \\ a & 0 & \epsilon \\ b
& \epsilon & 0
\end{array} \) = m_{\nu}^{T}
\label{genoff} \end{equation}
with $\det m_{\nu} = -2ab\epsilon$, and where we imagine the entry
$\epsilon$ to be a small perturbation around the basic structure
of~(\ref{lambdaT1}), which can arise from higher-order terms in
$W$. Without loss of generality we can redefine the phases of the
lepton doublets $L_i$ and $L'_i$ such that the entries $a$, $b$
and $\epsilon$ are real and $m_{\nu} = m_{\nu}^{\dagger}$. This
implies
\begin{equation}
U_{\nu}^{\dagger}m_{\nu} U_{\nu} = {\rm diag}(m_1, m_2, m_3)
\equiv m_{\rm diag} .
\label{Unu} \end{equation}
We also have $\Tr \; m_{\rm diag} = m_1 + m_2 + m_3 = \Tr \;
m_{\nu} = 0$, where the various eigenvalues $m_i$ are real but can
be negative.

If we begin by first ignoring the solar mass difference, and take
the atmospheric mass difference to be given by~(\ref{dm23}), then
there is no way to accommodate the ``normal'' hierarchy while
maintaining the requirement that $\epsilon \ll a,b$. For the
inverted hierarchy (in the same approximation of vanishing solar
mass difference) we would require $m_2 = -m_1 = 0.049 \; {\rm eV}$
with $m_3 =0$. This could derive from~(\ref{genoff}) if $\sqrt{a^2
+ b^2} = m_2$ and $\epsilon=0$. In this case $\sum_i |m_i| = 0.098
\; {\rm eV}$. This is clearly in line with the form of~(\ref{WT})
and implies a triplet mass of order
\begin{equation}
M_T = 2.0 \times \lambda_2 \lambda_T \(\frac{v_2 v'_2}{(100
\GeV)^2}\) \times 10^{14} \GeV
\label{MT} \end{equation}
where we have defined $v_2 = \lang (h_2)^0 \rang$ and $v'_2 =
\lang (h'_2)^0 \rang$. The solar mass difference~(\ref{dm12}) can
be restored in this case by taking $\epsilon \simeq \frac{1}{43}$
in the mass matrix given by
\begin{equation}
m_{\nu} = \sqrt{\frac{|\Delta m^2_{23}|}{2}} \(
\begin{array}{ccc} 0 & -1 & -1 \\ -1 & 0 & \epsilon
\\ -1 & \epsilon & 0 \end{array} \) .
\label{perturb} \end{equation}

This value is particularly encouraging for theories motivated by
the weakly coupled heterotic string compactified on orbifolds.
Such theories generally give rise to an Abelian gauge factor with
non-vanishing trace anomaly. This anomaly is cancelled by the
Green-Schwarz mechanism, which involves a Fayet-Iliopoulos (FI)
term $\xi_{\rm FI}$ in the 4D
Lagrangian~\cite{Dine:1987xk,Dine:1987gj,Atick:1987gy}. In
general, at least one field $X$ of the massless spectrum, charged
under this anomalous $U(1)$ factor, will receive a vev $X \simeq
\sqrt{\xi_{\rm FI}}$ so as to ensure $\lang D_X \rang = 0$ below
the scale $\xi_{\rm FI}$. Explicit orbifold constructions suggest
that $0.09 \leq r_{\rm FI} = \sqrt{|\xi_{\rm FI}|}/M_{\PL} \leq
0.14$ for $g^2 \simeq 1/2$~\cite{Giedt:2001zw}. Thus the
perturbation $\epsilon$ could be the result of non-renormalizable
operators in the superpotential of relative low-degree -- perhaps
involving only one or two powers of such a field vev, depending on
the size of the dimensionless Yukawa couplings involved.

Considering the underlying $SU(2)_A \times SU(2)_B$ theory, fields
bifundamental under both $SU(2)$ factors will decompose into a
triplet and a singlet under the breaking to the diagonal subgroup.
Let us denote this singlet representation as $\psi$. Then terms at
dimension four in the superpotential that can populate the
vanishing entries in~(\ref{lambdaT1}) include
\begin{eqnarray}
\Delta W &=& \frac{\lambda_{11}}{M_{\PL}} L_1 (\mathbf{2}, 1) T
(\mathbf{2}, \mathbf{2}) \psi (\mathbf{2}, \mathbf{2}) L_1
(\mathbf{2}, 1) \nonumber \\
 & & +\frac{\lambda_{ij}}{M_{\PL}}L'_i (1, \mathbf{2})
 T(\mathbf{2},\mathbf{2}) \psi (\mathbf{2},\mathbf{2}) L'_j
 (1,\mathbf{2}) ,
\label{psi} \end{eqnarray}
where $i, j = 2, 3$ and we denote the representations under
$SU(2)_A \times SU(2)_B$ for convenience. The singlet field $\psi$
must have vanishing hypercharge, so it cannot be the singlet
component of the same bifundamental that led to $T$ and
$\oline{T}$, though it may be the singlet component of some
bifundamental representation that served to generate the breaking
to the diagonal subgroup in the first place, or could be
identified with $\wh{S}_4$ or $\wh{S}_5$ of~(\ref{Wfull}). To the
extent that string models seldom give self-couplings at such a low
order in the superpotential, we might expect $\lambda_{11}  =
\lambda_{22} = \lambda_{33} =0$, thereby
generating~(\ref{perturb}) at roughly the correct order of
magnitude.

Now let us consider the leptonic (PMNS) mixing matrix  defined by
$U_{\rm PMNS} = U_e^{\dagger}U_{\nu}$, where $U_{\nu}$ is the
matrix in~(\ref{Unu}) and $U_e$ is the analogous matrix for the
charged leptons. Most of the earlier
studies~\cite{DeGouvea:2005gd,Langacker:2004xy,Altarelli:2004za,Mohapatra:2004vr,Barger:1998ta,Barbieri:1998mq,Altarelli:1998nx,Jezabek:1998du,Mohapatra:1999zr,Babu:2002ex}
of the texture in (\ref{lambdaT1}) assumed that this form holds in
the basis for which $U_e=1$. In that case, one has an inverted
hierarchy and $U_{\rm PMNS} = U_{\nu}$ is bimaximal for $a=b$,
i.e., $\theta_{12}=\theta_{23}=\pi/4$, while $\theta_{13}=0$. For
$|a|\ne |b|$ the solar mixing remains maximal while the
atmospheric mixing angle is $|\tan \theta_{23}|=|b|/|a|$. It is
now well established, however, that the solar mixing is not
maximal, i.e.,
$\pi/4-\theta_{12}=0.19^{+0.05}_{-0.06}$~\cite{Fogli:2005cq},
where the quoted errors are~2$\sigma$. It is well-known that
reasonable perturbations on this texture (still with $U_e=1$) have
difficulty yielding  a realistic solar mixing and mass splitting.
To see this, let us add a perturbation $\delta$ to the 13-entry of
(\ref{perturb}) and perturbations $\epsilon_{ii}$ to the diagonal
entries. To leading order, $\delta$ only shifts the atmospheric
mixing from maximal (to $\theta_{12}\sim \pi/4-\delta/2$).
$\epsilon_{22}$ and $\epsilon_{33}$ large enough to affect the
Solar mixing tend to give too large contributions to
$|\theta_{13}|$, so we will ignore them (their inclusion would
merely lead to additional fine-tuned parameter ranges). One then
finds
\begin{equation}
\frac{\pi}{4} - \theta_{12} \simeq 0.19 \simeq
\frac{1}{4}(\epsilon - \epsilon_{11}) ,
\label{solardev} \end{equation}
whereas the solar mass difference is
\begin{equation}
\frac{\Delta m^2_{12}}{\sqrt{2}|\Delta m_{\rm atm}^2|} \simeq
\frac{1}{43} \simeq (\epsilon + \epsilon_{11}) .
\label{ratio} \end{equation}
Satisfying these constraints would require a moderate tuning of
$\epsilon$ and $\epsilon_{11}$. Moreover, they would each have to
be of order~0.4 in magnitude, somewhat large to be considered
perturbations.

On the other hand, a simple and realistic pattern emerges when we
instead allow for small departures from $U_e \propto
\mathbf{1}$~\cite{Frampton:2004ud,Rodejohann:2003sc,Petcov:2004rk,Altarelli:2004jb,Romanino:2004ww,Datta:2005ci},
and for the general superpotential in (\ref{Wfull}) there is no
reason for such mixings to be absent.\footnote{The relatively
large value required for $\sin\theta_{12}^e$ compared to
$\sqrt{m_e/m_\mu}\sim 0.07$ suggests an asymmetric charged lepton
mass matrix, but this would not be unexpected.}  For example,
starting from~(\ref{perturb}) a Cabibbo-sized 12-entry in the
charged lepton mixing matrix
\begin{equation}
U_e^{\dagger} \sim \left( \begin{array}{ccc} 1 &
-\sin\theta_{12}^e & 0
\\ \sin\theta_{12}^e & 1 & 0 \\ 0 & 0 &1 \end{array}\right) ,
\end{equation}
leads to
\begin{equation}
\frac{\pi}{4} - \theta_{12} \simeq \frac{\sin\theta_{12}^e}{\sqrt{2}},%
\end{equation}
which is satisfied for $\sin\theta_{12}^e\simeq
0.27^{+0.07}_{-0.08}$. This mixing also leads to the prediction of
a large
\begin{equation}
\sin^2 \theta_{13}\simeq \frac{\sin^2\theta_{12}^e}{{2}}
\simeq(0.017-0.059)
\end{equation}
(the range is $\pm 2\sigma$), close to the current experimental
upper limit of 0.032. Finally, this model implies
\begin{equation}
m_{\beta\beta} \simeq m_2 (\cos^2 \theta_{12}-\sin^2
\theta_{12})\simeq 0.018 {\rm \ eV}
\end{equation}
for the effective mass relevant to neutrinoless double beta decay.
This is the standard result for the inverted hierarchy, with the
minus sign due to the opposite signs of $m_1$ and $m_2$. Such a
value should be observable in planned
experiments~\cite{DeGouvea:2005gd,Langacker:2004xy,Altarelli:2004za,Mohapatra:2004vr}.

\section{Realization in Heterotic String Models}
\label{sec:string}

Having outlined in a broad manner the elementary requirements for
phenomenological viability of any triplet-based model with a
structure dictated by the superpotential in~(\ref{Wfull}), we
might now wish to ask whether such a set of fields and couplings
really does arise in explicit string constructions as we have been
assuming. Rather than build all possible constructions of a
certain type for a dedicated scan -- an undertaking that would
undoubtedly produce interesting results in many areas, but which
we reserve for a future study -- we will here choose one
particular example as a case study. The $\mathbb{Z}_3 \times
\mathbb{Z}_3$ orbifold construction of Font et
al.~\cite{Font:1989aj} begins with a non-standard embedding that
utilizes two shift vectors and one Wilson line in the first
complex plane. This Wilson line breaks the observable sector gauge
group from $SO(10)$ to $SU(3) \times SU(2)_A \times SU(2)_B$. The
massless spectrum of this model contains~75 species of fields.
Those from the untwisted sectors have a multiplicity of one, while
twisted sectors have a multiplicity of three or nine, depending on
the representation. It is natural to consider this multiplicity
factor as a generation index.

There are three species of fields which are bifundamental under
the observable sector $SU(2)_A \times SU(2)_B$ (one in the
untwisted sector and two in various twisted sectors), five
doublets under $SU(2)_A$ and eight doublets under $SU(2)_B$. There
were also~17 species that were singlets under all non-Abelian
groups. So the minimal set of fields needed to generate the
superpotential of~(\ref{Wfull}) are present, as well as an
additional bifundamental representation that may be used to break
the product group to the diagonal subgroup and/or generate the
needed higher-order corrections in~(\ref{psi}). We note that there
are additional species that have non-trivial representations under
the non-Abelian groups of {\em both} the observable and hidden
sectors. In order to avoid potential complications should any of
these hidden sector groups undergo confinement we have not
considered these in what follows.

From the selection rules given in~\cite{Font:1989aj} it is
possible to construct all possible dimension three
(renormalizable) and dimension four (non-renormalizable)
superpotential couplings consistent with gauge invariance.
Considering only the~33 relevant fields mentioned in the previous
paragraph, the selection rules and gauge invariance under the
observable and hidden sector non-Abelian groups allow~32 and~135
terms at dimension three and four, respectively. Requiring in
addition gauge invariance under the six $U(1)$ factors (one of
which being anomalous) reduces these numbers to a tractable~15
and~8, respectively.

To ascertain which of the terms in~(\ref{Wfull}) can be identified
from the above it is necessary to choose a linear combination of
the five non-anomalous $U(1)$ factors to be identified as
hypercharge, and then determine the resulting hypercharges of the
bifundamentals, doublets and singlets under this assignment. Our
algorithm was to begin with the two bifundamentals of the twisted
sector, as the untwisted bifundamental had no couplings to $SU(2)$
doublets at dimension three or four. These two twisted sector
fields had a selection-rule allowed coupling to a non-Abelian
group singlet at the leading (dimension three) order, which could
therefore play the role of $S_3$ in~(\ref{Wfull}). Requiring these
two fields to carry hypercharge $Y= \pm 1$ (and thus automatically
ensuring that the candidate $S_3$ have vanishing hypercharge)
placed two constraints on the allowed hypercharge embedding.

We then proceed to the coupling~(\ref{WT}), or $\lambda_T$
in~(\ref{Wfull}) for the $Y=+1$ species. Each bifundamental had
several couplings of this form to various pairs of $SU(2)$
doublets, at both dimension three and dimension four. By
considering all possible pairs and requiring that the doublets
involved be assigned $Y= -1/2$ places two more constraints on the
allowed hypercharge embedding. Finally, we proceed to the equally
critical $\lambda_2$ coupling in~(\ref{Wfull}) for the oppositely
charged $Y=-1$ species. Again by considering all possible pairs of
doublets with this coupling, and requiring that both have
hypercharge $Y=+1/2$ we typically constrained the hypercharge
embedding to a unique embedding. The hypercharges of all the
states in the theory are then determined. Not all will have
Standard Model hypercharges, and thus most will have fractional or
non-standard electric charges and must be discarded as
``exotics.'' From the set with Standard Model hypercharge
assignments we can identify the surviving couplings
of~(\ref{Wfull}). In all, this process resulted in~35 distinct
field assignment possibilities, each having as a minimum the
couplings $\lambda_T$, $\lambda_2$ and $\lambda_3$ -- the minimum
set to generate the triplet see-saw and the mass pattern
of~(\ref{lambdaT1}). Though these couplings are not enough to
generate the perturbations on the bimaximal texture, nor do they
include the couplings needed to generate charged lepton masses or
$\mu$-terms to break electroweak symmetry, they still represent a
complete set of needed couplings to explain the smallness of
neutrino masses generally -- something that a more exhaustive
search of a whole class for the ``standard'' seesaw failed to
achieve~\cite{Giedt:2005vx}.

None of the~35 possibilities allowed for all of the couplings
of~(\ref{Wfull}), and~12 had no other couplings than the minimal
set. This is yet another example of how selection rules of the
underlying conformal field theory often forbid operators that
would otherwise be allowed by gauge invariance in the 4D~theory.
Rather than present the various features of all of these
assignments, we instead point out a few particular cases. One
successful hypercharge assignment allows for a superpotential of
the form
\begin{eqnarray}
W &=& \lambda_T L T L' + \lambda_2 H_2 \oline{T} H'_2 + \lambda_3
S_3 T \oline{T} \nonumber \\
& & + \lambda_5 S_5 H'_1 H'_2 + \lambda_7 E_R L' H'_1 ,
\label{W2} \end{eqnarray}
where in this case $L$, $L'$, $H_2$ and $H'_2$ all have
multiplicity three, $H'_1$ has multiplicity one and there is no
species with the correct hypercharge to be identified as $H_1$. In
this case, identifying $L$ with the doublet containing the
electron leaves the electron massless after electroweak symmetry
breaking up to terms of dimension five in the superpotential.

Alternatively, one can obtain candidates for all six species of
doublets, such as an example in which the allowed superpotential
is given by
\begin{eqnarray}
W &=& \lambda_T L T L' + \lambda_1 H_1 T H'_1 + \lambda_2 H_2
\oline{T} H'_2 \nonumber
\\
& & + \lambda_3 S_3 T \oline{T} + \lambda_7 E_R L' H'_1 .
\label{W3} \end{eqnarray}
All doublets except $H_2$ in this case arise from twisted sectors,
so have multiplicity three. It is interesting to note that in
several of the~35 cases the hypercharge embedding assigned $Y=0$
to the bifundamental representation of the untwisted sector,
suggesting it could play the role of breaking the product group to
the diagonal subgroup. Couplings of the form of~(\ref{psi}),
however, were forbidden by the string theoretic selection rules
through dimension four.

Of course none of these cases are truly realistic in the sense of
what is needed to explain the observed neutrino data as outlined
in the previous section, and it would have been naive to have
expected any to be in the first place. The above examples are
instead meant to demonstrate the plausibility of this new
realization of a triplet-induced seesaw from a string-theory
viewpoint by means of a ready example from the literature. Having
introduced the concept, defined a basic structure as
in~(\ref{Wfull}) and demonstrated that the structure may in fact
be realized in the context of tractable string constructions, it
becomes reasonable to propose a dedicated search over a wide class
of constructions for precisely this model -- a search that would
necessarily be a separate research project in its own right but
which would complement well the analysis already performed
in~\cite{Giedt:2005vx}.

The $\mathbb{Z}_3 \times \mathbb{Z}_3$ construction is often
considered because it, like its $\mathbb{Z}_3$ cousin, generates a
three-fold redundancy for most of the massless spectrum in a
relatively straightforward way. But a minimal model would
presumably prefer to break away from the three-fold degeneracy on
every species, but not the requirement of three generations
globally. For example, it is possible to imagine a model in which
there are only three ``lepton'' doublets of $SU(2)_L$ once we
break to the diagonal subgroup. Since species in orbifold models
(and orientifold models of open strings as well) are defined by
fixed point locations (i.e., geometrically) this is not
unreasonable to imagine -- in fact, precisely such a separation of
the three ``generations'' occurs in the recent $\mathbb{Z}_2\times
\mathbb{Z}_3$ construction of Kobayashi et
al.~\cite{Kobayashi:2004ud}. Nevertheless, there is no getting
around the need for at least an extra pair of one, if not both, of
the Higgs doublets of the MSSM. As discussed
following~(\ref{Wfull}), it is possible that the extra doublets
are projected out near the string scale (e.g., if some of the
$S_{4,5}$ and $\wh{S}_{4,5}$ are associated with the
Fayet-Iliopoulos terms) or at an intermediate scale. It is also
possible that one or more extra doublets survives to the TeV
scale, in which case there are potential implications for
FCNC~\cite{Sher:1991km,Atwood:1996vj} and CP
violation~\cite{Hall:1993ca}, as well as for the charged lepton
mixing generated from~(\ref{Wfull}). A more detailed study of such
issues is beyond the scope of this paper.

\section*{Conclusions}

We have presented a new construction of Type~II seesaw models
utilizing triplets of $SU(2)_L$ in which that group is realized as
the diagonal subgroup of an $SU(2)_A \times SU(2)_B$ product
group. The triplets in this construction begin as bifundamentals
under the two original $SU(2)$ factors, and this identification
immediately leads to a bimaximal mixing texture for the effective
neutrino mass matrix provided generations of lepton doublets are
assigned to the two underlying $SU(2)$ factors in the appropriate
way. The observed atmospheric mass difference can be accommodated
if the triplets obtain a mass of order $10^{14} \GeV$, and the
solar mass difference can easily be incorporated by a simple
perturbation arising at dimension four or five in the
superpotential. The observed deviation of the Solar mixing from
maximal can be accommodated by a small (Cabibbo-like) mixing in
the charged lepton sector, leading to predictions for
$\theta_{13}$ and neutrinoless double beta decay.

We were led to consider this construction by imagining the
simplest possible requirements for generating a triplet of
$SU(2)_L$ from string constructions -- particularly the weakly
coupled heterotic string, though the model can be realized in
other constructions as well. Though inspired by string theory, the
model is not itself inherently stringy and is interesting in its
own right. Some of the properties of this model are known to
phenomenologists, who have arrived at a similar mass matrix from
other directions. Interestingly, however, to the best of our
knowledge the particular texture has not emerged from other
versions of heavy triplet models, e.g., motivated by grand
unification or left-right symmetry. The simplest version of the
construction requires at least one additional pair of Higgs
doublets, which may or may not survive to the TeV scale.

Having laid out a concrete model as a plausible alternative to the
standard Type~I seesaw in string-based constructions, it is now
possible to examine large classes of explicit string models to
search for both types of neutrino mass patterns. Given the
difficulty in finding a working example of the minimal Type~I
seesaw in at least one otherwise promising class of string
construction, having alternatives with clear ``signatures'' (in
this case, at least two $SU(2)$ factors, with at least two fields
bifundamental under both, capable of forming a hypercharge-neutral
mass term) is welcome.

\begin{acknowledgments}
We wish to thank Joel Giedt and Boris Kayser for helpful
discussions and advice. This work was supported by the
U.S.~Department of Energy under Grant No.~DOE-EY-76-02-3071.
\end{acknowledgments}


\begin{thebibliography}{99}
\bibitem{Maltoni:2003da}
  M.~Maltoni, T.~Schwetz, M.~A.~Tortola and J.~W.~F.~Valle,
  Phys.\ Rev.\ D {\bf 68}, 113010 (2003).
\bibitem{Bahcall:2004ut}
  J.~N.~Bahcall, M.~C.~Gonzalez-Garcia and C.~Pena-Garay,
  JHEP {\bf 0408}, 016 (2004).
\bibitem{Fogli:2005cq}
  G.~L.~Fogli, E.~Lisi, A.~Marrone and A.~Palazzo,
  ``Global analysis of three-flavor neutrino masses and mixings,''
  [arXiv:hep-ph/0506083].

\bibitem{King:2003jb}
  S.~F.~King,
  Rept.\ Prog.\ Phys.\  {\bf 67}, 107 (2004).
\bibitem{DeGouvea:2005gd}
  A.~De Gouvea,
  Mod.\ Phys.\ Lett.\ A {\bf 19}, 2799 (2004).
\bibitem{Langacker:2004xy}
  P.~Langacker, ``Neutrino physics (theory),''
  [hep-ph/0411116].
\bibitem{Altarelli:2004za}
  G.~Altarelli and F.~Feruglio,
  New J.\ Phys.\  {\bf 6}, 106 (2004).
\bibitem{Mohapatra:2004vr}
  R.~N.~Mohapatra {\it et al.},
  ``Theory of neutrinos,''
  [arXiv:hep-ph/0412099].

\bibitem{Giedt:2005vx}
  J.~Giedt, G.~L.~Kane, P.~Langacker and B.~D.~Nelson,
  ``Massive neutrinos and (heterotic) string theory,''
  [hep-th/0502032].

\bibitem{Lazarides:1980nt}
  G.~Lazarides, Q.~Shafi and C.~Wetterich,
  Nucl.\ Phys.\ B {\bf 181}, 287 (1981).
\bibitem{Mohapatra:1980yp}
  R.~N.~Mohapatra and G.~Senjanovic,
  Phys.\ Rev.\ D {\bf 23}, 165 (1981).
\bibitem{Schechter:1981cv}
  J.~Schechter and J.~W.~F.~Valle,
  Phys.\ Rev.\ D {\bf 25}, 774 (1982).

\bibitem{Ma:1998dx}
  E.~Ma and U.~Sarkar,
  Phys.\ Rev.\ Lett.\  {\bf 80}, 5716 (1998).
\bibitem{Hambye:2000ui}
  T.~Hambye, E.~Ma and U.~Sarkar,
  Nucl.\ Phys.\ B {\bf 602}, 23 (2001).
\bibitem{Rossi:2002zb}
  A.~Rossi,
  Phys.\ Rev.\ D {\bf 66}, 075003 (2002).

\bibitem{Font:1989aj}
  A.~Font, L.~E.~Ibanez, F.~Quevedo and A.~Sierra,
  Nucl.\ Phys.\ B {\bf 331}, 421 (1990).

\bibitem{GRS}
  M.~Gell-Mann, P.~Ramond and R.~Slansky in {\it Supergravity},
  ed. P.~van~Nieuwenhuizen and D.~Z.~Freedman (North-Holland,
  Amsterdam), 1979.
\bibitem{Yanagida}
  T.~Yanagida in {\it Proceedings of the Workshop on Unified
  Theory and Baryon Number in the Universe}, ed. O.~Sawada and
  A.~Sugamoto, KEK Report KEK-97-18, 1979.
\bibitem{Valle1}
  J.~Schechter and J.~W.~F.~Valle, Phys.\ Rev.\ D {\bf 22}, 2227
  (1980).


\bibitem{Gross:1985fr}
  D.~J.~Gross, J.~A.~Harvey, E.~J.~Martinec and R.~Rohm,
  Nucl.\ Phys.\ B {\bf 256}, 253 (1985).
\bibitem{Font:1990uw}
  A.~Font, L.~E.~Ibanez and F.~Quevedo,
  Nucl.\ Phys.\ B {\bf 345}, 389 (1990).

\bibitem{Dienes:1996yh}
  K.~R.~Dienes and J.~March-Russell,
  Nucl.\ Phys.\ B {\bf 479}, 113 (1996).

\bibitem{Barbieri:1994jq}
  R.~Barbieri, G.~R.~Dvali and A.~Strumia,
  Phys.\ Lett.\ B {\bf 333}, 79 (1994).

\bibitem{Barger:1998ta}
  V.~D.~Barger, S.~Pakvasa, T.~J.~Weiler and K.~Whisnant,
  Phys.\ Lett.\ B {\bf 437}, 107 (1998).
\bibitem{Barbieri:1998mq}
  R.~Barbieri, L.~J.~Hall, D.~R.~Smith, A.~Strumia and N.~Weiner,
  JHEP {\bf 9812}, 017 (1998).
\bibitem{Altarelli:1998nx}
  G.~Altarelli and F.~Feruglio,
  Phys.\ Lett.\ B {\bf 439}, 112 (1998).
\bibitem{Jezabek:1998du}
  M.~Jezabek and Y.~Sumino,
  Phys.\ Lett.\ B {\bf 440}, 327 (1998).

\bibitem{Mohapatra:1999zr}
  R.~N.~Mohapatra, A.~Perez-Lorenzana and C.~A.~de Sousa Pires,
  Phys.\ Lett.\ B {\bf 474}, 355 (2000).
\bibitem{Babu:2002ex}
  K.~S.~Babu and R.~N.~Mohapatra,
  Phys.\ Lett.\ B {\bf 532}, 77 (2002).

\bibitem{Giudice:1988yz}
  G.~F.~Giudice and A.~Masiero,
  Phys.\ Lett.\ B {\bf 206}, 480 (1988).


\bibitem{Dine:1987xk}
  M.~Dine, N.~Seiberg and E.~Witten,
  Nucl.\ Phys.\ B {\bf 289}, 589 (1987).
\bibitem{Dine:1987gj}
  M.~Dine, I.~Ichinose and N.~Seiberg,
  Nucl.\ Phys.\ B {\bf 293}, 253 (1987).
\bibitem{Atick:1987gy}
  J.~J.~Atick, L.~J.~Dixon and A.~Sen,
  Nucl.\ Phys.\ B {\bf 292}, 109 (1987).
\bibitem{Giedt:2001zw}
  J.~Giedt,
  Annals Phys.\  {\bf 297}, 67 (2002).

\bibitem{Frampton:2004ud}
  P.~H.~Frampton, S.~T.~Petcov and W.~Rodejohann,
  Nucl.\ Phys.\ B {\bf 687}, 31 (2004).

\bibitem{Rodejohann:2003sc}
  W.~Rodejohann,
  Phys.\ Rev.\ D {\bf 69}, 033005 (2004).
\bibitem{Petcov:2004rk}
  S.~T.~Petcov and W.~Rodejohann,
  Phys.\ Rev.\ D {\bf 71}, 073002 (2005).
\bibitem{Altarelli:2004jb}
  G.~Altarelli, F.~Feruglio and I.~Masina,
  Nucl.\ Phys.\ B {\bf 689}, 157 (2004).
\bibitem{Romanino:2004ww}
  A.~Romanino,
  Phys.\ Rev.\ D {\bf 70}, 013003 (2004).
\bibitem{Datta:2005ci}
  A.~Datta, L.~Everett and P.~Ramond,
  ``Cabibbo haze in lepton mixing,''
  [arXiv:hep-ph/0503222].



\bibitem{Kobayashi:2004ud}
  T.~Kobayashi, S.~Raby and R.~J.~Zhang,
  Phys.\ Lett.\ B {\bf 593}, 262 (2004).

\bibitem{Sher:1991km}
  M.~Sher and Y.~Yuan,
  Phys.\ Rev.\ D {\bf 44}, 1461 (1991).
\bibitem{Atwood:1996vj}
  D.~Atwood, L.~Reina and A.~Soni,
  Phys.\ Rev.\ D {\bf 55}, 3156 (1997).
\bibitem{Hall:1993ca}
  L.~J.~Hall and S.~Weinberg,
  Phys.\ Rev.\ D {\bf 48}, 979 (1993).

\end{thebibliography}

\end{document}